\documentclass{article}
\usepackage[latin9]{inputenc}
\usepackage{amsfonts}
\usepackage{amsmath}

\input{tcilatex}

\begin{document}

\begin{titlepage}
\vspace{1cm}
\baselineskip=24pt
\begin{center}
\textbf{\LARGE{ Generalization of extended Lie algebras by
expansions of extended de Sitter algebra, in four dimensions.}}
\par \end{center}{\LARGE \par}

\begin{center}
	\vspace{1cm}
	\textbf{Ricardo Caroca}

   \textit{Departamento de Matemática y Física Aplicadas, }\vspace{-8pt}
	\textit{ Universidad Católica de la Santísima Concepción, }\vspace{-8pt}

                \textit{ Alonso de Ribera 2850, Concepción, Chile.}
	\\[1mm]
\footnotesize
	\texttt{rcaroca@uscs.cl}
	\end{center}

\vskip 26pt
\begin{abstract}
\noindent
Four-dimensional extended: Poincaré, AdS-Lorentz and Maxwell algebras, are obtained by expanding an extension of de Sitter or conformal algebra, SO(4,1) or SO(3,2). The procedure can be generalized to obtain a new family of extended $\mathcal{C}_k^{E}$ and its flat limit, the extended $\mathcal{B}_k^{E}$ algebras. The  extended $\mathcal{C}_k$ and $\mathcal{B}_k$ algebras have been introduced in the literature recently.The extended Poincaré algebra is also obtained as an Inönü-Wigner contraction of extended de Sitter algebra.

\end{abstract}
\end{titlepage}

\section{Introduction}

\ \ \ \ One of the great challenges of contemporary physics is to reconcile
quantum mechanics with general relativity, the two last and greatest
conceptual revolutions of physics, to describe the four fundamental
interactions in a single theoretical framework, in addition to solving a
series of phenomenological problems that exist today, it's still an open
problem \cite{Rovelli2004}.

The supersymmetry and the supergravity theories may remove some ultraviolet
divergences in perturbation theory, which is one of the big problems today,
but the problem with supergravity theory is that the restriction $s\leq 2$
on the spins of the particles in the massless supergravity supermultiplet
leads to a famous limitation $N\leq 8$ on the number of gravitinos. The
gravitino is a massless spin-two field and it is the quantum mechanical
particle that carries gravitational force and therefore plays a crucial
role, showing the inability of the supergravity to work with interacting
higher-spin gauge fields \cite{V1996}. \ For its part, the string theories
that involve fields of arbitrary spin (higher-spin), the ultraviolet
divergence is solved because the point-like particles of the Standard Model
are replaced by extended objects called strings. Fundamental properties such
as: mass, charge and other, are determined by the vibrational state of the
string, and one of the vibrational states gives rise to the graviton.
Therefore, the string theory incorporates all of the fundamental
interactions, including gravity and the superstring theory describe infinite
collections of higher-spin excitations of all massive spins \cite{GSW}, \cite%
{EW1995}. \ Five consistent versions ended being different limits of the
M-theory in eleven dimensions. Lamentably, there are no clear observable
predictions in the experiments, due to the high energies they require,
besides having some unsolved problems such as the compactified dimensions
and some mathematical problems. Therefore, it is natural to ask the
following question: is there a certain spontaneous symmetry breaking
mechanism from some underlies symmetric phase \cite{V1996}, to explain the
massive excitations of string theories ?. There is gauge theories of
interacting fields of all spin that are based in infinite-dimensional gauge
symmetries, and its are know as higher-spin gauge theories (HS) \cite{V1996}%
, \cite{FV1987}-\cite{VHS}, which can be considered as most general gauge
theories in (3+1) space-time dimensions. The higher-spin gauge symmetries
describe infinite collection of interacting higher-spin massless field of
all spin $0\leq s\leq \infty $ \cite{FV1987}-\cite{VHS} and notably the HS
theories contain lower-spin gauge symmetries ($s\leq 2$) as
finite-dimensional subalgebras \cite{V1996}.

On the other hand, a higher-spin gauge extension coupled to the
three-dimensional AdS gravity was formulated as a Chern-Simons theory \cite%
{CFPT} whose gauge group is given by $SL(n)\times SL(n)$ and two copies of $%
\mathcal{W}$-algebras as asymptotic symmetries of asymptotically Anti-de
Sitter solutions was obtained. \ The particular case of a spin-3 field for
the group $SL(3)\times SL(3$) was studied in detail in Ref. \cite{CFPT} and
two copies of the classical $\mathcal{W}_{3}$-algebra with central charge
was resulted, describing the coupling of a spin-3 gauge field to gravity.

Another interesting point of view is to extend the Poincaré group,
introdu-cing fermionic generators with half-integer spin generators, in
three space-time dimensions. It was shown that hypergravity could be
formulated as a local gauge symmetry \cite{FMT2015}, \cite{FMT2}.\ An
Chern-Simons action invariant to the extension of the Poincaré group can be
formulated, since this algebra admits a nontrivial Casimir operator. An
hypergravity in the generic case of massless fermionic field, could be
formulated, to the case of hyper-Poincaré algebra.

Another way to approach to the higher-spin gauge symmetries, is the proposal
to extend the (Super)Poincaré group, and consists of introducing a mix of
internal and space-time symmetries \cite{SG}, \cite{Antoniadis2011}, \cite%
{Antoniadis2012}, in four space-time dimensions. Starting from the base
generators of an internal compact Lie algebra, it consists of introducing an
infinite set of translationally invariant generators, which are totally
symmetrical with respect to the space-time indices. These generators close
as subalgebra in the extended Poincaré algebra, are s-range tensors and have
a nonzero commutation relation with the Lorentz generators, and therefore
are higher spin carriers. The massless case and the irreducible
representations of the extended symmetry were constructed. The transversal
representations imply an infinite series of helicities of integers and
semi-integers.

An extension of the de Sitter and conformal group with the same approach is
suggested, and will be the starting point of this work \cite{Antoniadis2011}-%
\cite{Savvidy2006}. We can say that the construction of consistent
"higher-spin gauge theory", suggests that the algebra that supports it must
be infinite-dimensional. In what follows, we will use the classification
which have been recently introduced in the literature in the context of:
(super)gravity \cite{P. Salgado2014}, \cite{CR1}, higher-spin gravity \cite%
{CCFRS} and the asymptotic symmetries \cite{CCRS}; to designate the
generalizations of the families algebras: the AdS ($\mathcal{C}_{3}$) and
Poincaré ($\mathcal{B}_{3}$), AdS$\oplus $Lorentz ($\mathcal{C}_{4}$) and
Maxwell ($\mathcal{B}_{4}$), and therefore the generaliced $\mathcal{C}_{k}$
and $\mathcal{B}_{k}$ algebras \cite{Concha2013}, \cite{P. Concha2014}, \cite%
{CPRS3}.

The paper is organized as follows: In the section 2 we briefly review an
extension of de Sitter and conformal algebras ($\mathcal{C}_{3}^{E}$) and
subsequently, in section 2.1, we show that extended Poincaré algebra ($%
\mathcal{B}_{3}^{E}$) can be obtained as a Inönü-Wigner contraction of
extended de Sitter ($\mathcal{C}_{3}^{E})$ algebra. The section 3 is devoted
entirely to the results obtained by applying the S-expansion method to
extended de Sitter algebra ($\mathcal{C}_{3}^{E}$). In chapter 3.1, the
extended Poincaré Algebra ($\mathcal{B}_{3}^{E}$) as a resonant and reduced
subalgebra of $\mathcal{C}_{3}^{E}$, was obtained. Chapter 3.2 we construct
the commutations relations of extended AdS$\oplus $Lorentz ($\mathcal{C}%
_{4}^{E}$) through a reduced subalgebra of extended de Sitter algebra ($%
\mathcal{C}_{3}^{E}$). Subsequently, in section 3.3 by extending a flat
limit we rescued the extended Maxwell algebra ($\mathcal{B}_{4}^{E}$) from
extended AdS$\oplus $Lorentz. In the section 3.4 the previous results are
generalized, where extended $\mathcal{C}_{k}^{E}$ algebra as a reduced
subalgebra of extended de Sitter ($\mathcal{C}_{3}^{E}$) was obtained. In
the section 3.4 we found the extended $\mathcal{B}_{k}^{E}$ algebra as a
flat limit of extended $\mathcal{C}_{k}^{E}$ algebra. Finally, we conclude
with some comments and possible future developments.

\section{An extension of de Sitter and conformal algebras ($\mathcal{C}%
_{3}^{E}$).}

\ \ \ Consider first the de Sitter and conformal algebras $SO(4,1)$ or $%
SO(3,2)$, proposed by \cite{Antoniadis2011} (section VI, equation (39)),
whose generators satisfy the commutation relation%
\begin{equation}
\left[ J^{AB},J^{CD}\right]
=i(g^{AD}J^{BC}-g^{AC}J^{BD}+g^{BC}J^{AD}-g^{BD}J^{AC})  \label{Savv}
\end{equation}%
where $g^{AB}=\left( +----\right) $ or $g^{AB}=\left( +---+\right) $ \ and $%
A,B=0,1,..,4$. \ 

According to to \cite{Antoniadis2011}, \cite{Antoniadis2012}, \cite%
{Antoniadis2014}, \cite{KS2014}, \cite{Savvidy2005},\cite{Savvidy2006} we
might postulate an extension for the\ de Sitter and conformal groups. \ For
this, consider the extended de Sitter algebra in 4-dimensions, $\mathcal{A}d%
\mathcal{S}^{E}$, whose generators are%
\begin{equation*}
AdS^{E}=Span\left \{ J^{AB},L_{a}^{C_{1}...C_{s}}\right \} ,
\end{equation*}%
and the sets of commutators are the following \cite{Antoniadis2011}.

\begin{equation}
\left[ J^{AB},J^{CD}\right]
=i(g^{AD}J^{BC}-g^{AC}J^{BD}+g^{BC}J^{AD}-g^{BD}J^{AC})  \label{SLE1}
\end{equation}%
\begin{equation}
\left[ J^{AB},L_{a}^{C_{1}...C_{s}}\right] =i(\eta
^{C_{1}B}L_{a}^{AC_{2}...C_{s}}-...-\eta ^{C_{s}A}L_{a}^{C_{1}...C_{s-1}B})
\label{SLE2}
\end{equation}

\begin{equation}
\left[ L_{a}^{C_{1}...C_{n}},L_{b}^{C_{n+1}...C_{s}}\right]
=if_{abc}L_{c}^{C_{1}...C_{s}}\left( s=0,1,2,....\right) ,  \label{SLE4}
\end{equation}%
where the infinite set of generators $L_{a}^{\lambda _{1}...\lambda _{s}}$
is defined, in the form 
\begin{equation}
\begin{array}{ccc}
L_{a}^{\lambda _{1}...\lambda _{s}}=e^{\lambda _{1}}....e^{\lambda
_{s}}\otimes L_{a} & , & s=0,1,2,...%
\end{array}
\label{IG}
\end{equation}

The generators $L_{a}^{\lambda _{1}...\lambda _{s}}$ carry internal and
space-time indices, and are totally symmetric with respect to the indices $%
\lambda _{1}...\lambda _{s}$ \cite{Antoniadis2011}. The algebra (\ref{SLE1}%
)-(\ref{SLE4}) corresponds to a generalization to the cases of the de Sitter
groups.

On the other hand, the generators $L_{a}$ correspond to the internal compact
Lie algebra $L_{G}$, 
\begin{equation}
\begin{array}{ccc}
\left[ L_{a},L_{b}\right] =if_{abc}L_{c} & , & a,b,c=1,...,\dim L_{G}.%
\end{array}
\label{AIC}
\end{equation}%
where the internal bosonic algebra obeys the Coleman-Mandula theorem.

We can check that all Jacoby identities are satisfied and we have an
consistent algebra (\ref{SLE1})-(\ref{SLE4}).

\subsection{Extended Poincaré Algebra ($\mathcal{B}_{3}^{E}$) as a Inönü%
-Wigner contraction of extended de Sitter ($\mathcal{C}_{3}^{E})$ algebra.}

\ \ \ We rewrite the generators of $\mathcal{C}_{3}^{E}$ algebra (\ref{SLE1}%
)-(\ref{SLE4}) fixing some indices as follows: $\ $%
\begin{equation}
\begin{array}{cccc}
J^{4\mu }:=P^{\mu } & , & M^{\mu \nu }:=J^{\mu \nu } & ,%
\end{array}
\label{rg}
\end{equation}%
where $\mu ,\nu ,\rho =0,1,2,3$.\ \ Then, the extended (anti) Sitter
algebra, $\mathcal{A}d\mathcal{S}^{E}$( $SO(3,2)$ ) in 5-dimensions, written
in terms of the new generators, 
\begin{equation}
\mathcal{C}_{3}^{E}=Span\left \{ P^{\mu },J^{\mu \nu
},L_{a}^{C_{1}...C_{s}}\right \}  \label{AdSE}
\end{equation}%
where $sig(g^{\mu \nu })=\left( +---\right) $, has the following
commutations relations:%
\begin{eqnarray}
\left[ P^{\mu },P^{\nu }\right] &=&-iJ^{\mu \nu },  \label{AdSE1} \\
\left[ J^{\mu \nu },J^{\rho \sigma }\right] &=&i(g^{\mu \sigma }J^{\nu \rho
}-g^{\mu \rho }J^{\nu \sigma }+g^{\nu \rho }J^{\mu \sigma }-g^{\nu \sigma
}J^{\mu \rho }),  \label{AdSE2} \\
\left[ J^{\nu \rho },P^{\mu }\right] &=&i(g^{\mu \rho }P^{\nu }-g^{\mu \nu
}P^{\rho }),  \label{AdSE3}
\end{eqnarray}%
\begin{equation}
\left[ J^{\mu \nu },L_{a}^{C_{1}...C_{s}}\right] =i(\eta ^{C_{1}\nu
}L_{a}^{\mu C_{2}...C_{s}}-...-\eta ^{C_{s}\mu }L_{a}^{C_{1}...C_{s-1}\nu }),
\label{AdSE4}
\end{equation}%
\begin{equation}
\left[ P^{\mu },L_{a}^{C_{1}...C_{s}}\right] =i(\eta ^{C_{1}\mu
}L_{a}^{4C_{2}...C_{s}}-...-\eta ^{C_{s}4}L_{a}^{C_{1}...C_{s-1}\mu }),
\label{AdSE5}
\end{equation}

\begin{equation}
\left[ L_{a}^{C_{1}...C_{n}},L_{b}^{C_{n+1}...C_{s}}\right]
=if_{abc}L_{c}^{C_{1}...C_{s}}\left( s=0,1,2,....\right) .  \label{AdSE6}
\end{equation}

The $\mathcal{C}_{3}^{E}$ algebra (\ref{AdSE1})-(\ref{AdSE6}) presents a
structure of subalgebra $V_{0}$ and symmetric coset $V_{1}$ (\ref{V0V1-2});
where

\begin{equation}
\begin{array}{ccc}
V_{0}=\left \{ J^{\mu \nu },L_{a}^{C_{1}...C_{s}}\right \} & and & 
V_{1}=\left \{ P^{\mu }\right \}%
\end{array}
\label{V0V1}
\end{equation}

Making a Inönü-Wigner contraction \cite{IW}, \cite{IW2}, \cite{WW1}, \cite%
{WW2}: $P^{\mu }\rightarrow lP^{\mu }$andand extending the flat limit $%
l\longrightarrow \infty $, reduces it to the Extended Poincaré algebra ($%
\mathcal{B}_{3}^{E}$); that is to say: 
\begin{eqnarray}
\left[ \mathcal{P}^{\mu },\mathcal{P}^{\nu }\right] &=&0,  \label{PP} \\
\left[ \mathcal{J}^{\mu \nu },\mathcal{P}^{\rho }\right] &=&i(g^{\rho \nu }%
\mathcal{P}^{\mu }-g^{\rho \mu }\mathcal{P}^{\nu }),  \label{JP} \\
\left[ \mathcal{J}^{\mu \nu },\mathcal{J}^{\rho \sigma }\right] &=&i(g^{\mu
\sigma }\mathcal{J}^{\nu \rho }-g^{\mu \rho }\mathcal{J}^{\nu \sigma
}+g^{\nu \rho }\mathcal{J}^{\mu \sigma }-g^{\nu \sigma }\mathcal{J}^{\mu
\rho }),  \label{JJ} \\
\left[ \mathcal{P}^{\mu },\mathcal{L}_{a}^{C_{1}...C_{s}}\right] &=&0
\label{PL} \\
\left[ \mathcal{J}^{\mu \nu },\mathcal{L}_{a}^{C_{1}...C_{s}}\right]
&=&i(\eta ^{C_{1}\nu }\mathcal{L}_{a}^{\mu C_{2}...C_{s}}-...-\eta
^{C_{s}\mu }\mathcal{L}_{a}^{C_{1}...C_{s-1}\nu }),  \label{JL} \\
\left[ \mathcal{L}_{a}^{C_{1}...C_{n}},\mathcal{L}_{b}^{C_{n+1}...C_{s}}%
\right] &=&if_{abc}\mathcal{L}_{c}^{C_{1}...C_{s}}\left( s=0,1,2,....\right)
.  \label{LL}
\end{eqnarray}

Note that because the generators $\mathcal{L}_{a}^{C_{1}...C_{n}}$ have a
nonzero commutations relation with the generator $\mathcal{J}^{\mu \nu }$,
they carry higher spins gauge fields. In this case, the infinite set of
generators $L_{a}^{\lambda _{1}...\lambda _{s}}$ are translationally
invariant (\ref{PL}).

The first three commutations relations correspond to the usual Poincaré
algebra ($\mathcal{B}_{3}$) generated by $\left \{ \mathcal{P}^{\mu },%
\mathcal{J}^{\mu \nu }\right \} $, while the last three commutations
relations describe the coupling of the higher-spin generators $\left \{ 
\mathcal{L}_{a}^{C_{1}...C_{n}}\right \} $ to the Poincaré symmetry.

This last result (\ref{PP})-(\ref{LL}) corresponds the the same result
obtained in (9), (10) and (11) of reference \cite{Antoniadis2011}.

\section{S-expansions of extended de Sitter algebra ($\mathcal{C}_{3}^{E}$).}

\ \ \ In this section we will use the S-expansion procedure of Lie algebras 
\cite{Sexp}, \cite{IPRS}, which is briefly shown in Appendices A and B, to
obtain: extended Poincaré, extended AdS$\oplus $Lorentz, extended Maxwell,
and generalized extended $\mathcal{C}_{k}^{E}$ and generalized extended $%
\mathcal{B}_{k}^{E}$ algebras. The $C_{k}$ is a semisimple Lie algebras (AdS
"type") being a generalization that contains the AdS and AdS$\oplus $Lorentz
algebras, as particular cases. In contrast, $B_{k}$ is a nonsemisimple Lie
algebras (Poincaré "type") and are generalizations that contains the Poincaré
and Maxwell algebras as particular cases. It is important to note also that
all Jacobi identities (\ref{JC}) are guaranteed for expanded algebras due to
the associative and commutative property of the semigroup in the S-expansion
procedure, \cite{Sexp}, \cite{IPRS}. Therefore, all the expanded Lie
algebras are closed algebras, and have explicit matrix representations.

\subsection{Extended Poincaré Algebra ($\mathcal{B}_{3}^{E}$) as a resonant
and reduced subalgebra of $\mathcal{C}_{3}^{E}$.}

\ \ \ Using the S-expansion method of Lie algebras \cite{Sexp}-\cite{PDE},
shown in summary form in the appendices A and B, and let us consider the set 
$\ S^{(3)}=\left \{ \lambda _{0},\lambda _{1},\lambda _{2}\right \} $, with
the resonant partition (\ref{CR}) 
\begin{equation}
\begin{array}{ccc}
S_{0}=\left \{ \lambda _{0},\lambda _{2}\right \} & , & S_{1}=\left \{
\lambda _{1},\lambda _{2}\right \} ,%
\end{array}
\label{S0S1a}
\end{equation}%
to the resonant conditions (\ref{V0V1-2}) to the subspaces (\ref{V0V1}).
There are $2^{4}=16$ different sets that satisfy the resonant condition (\ref%
{V0V1-2}), and these have the following multiplication rules

\begin{equation}
\begin{array}{cccc}
\ast & \lambda _{0} & \lambda _{1} & \lambda _{2} \\ 
\lambda _{0} & \left \{ \lambda _{0},\lambda _{2}\right \} & \left \{
\lambda _{1},\lambda _{2}\right \} & \lambda _{2} \\ 
\lambda _{1} & \left \{ \lambda _{1},\lambda _{2}\right \} & \left \{
\lambda _{0},\lambda _{2}\right \} & \lambda _{2} \\ 
\lambda _{2} & \lambda _{2} & \lambda _{2} & \lambda _{2}%
\end{array}
\label{CSR}
\end{equation}%
and for example, the set $\left \{ \lambda _{0},\lambda _{2}\right \} $ in
the table means the product $\lambda _{0}\ast \lambda _{0}$ can be $\lambda
_{0}$ or $\lambda _{2}.$ From the multiplication table (\ref{CSR}), we
choose the following semigroup 
\begin{equation}
S^{(3)}=%
\begin{array}{cccc}
\ast & \lambda _{0} & \lambda _{1} & \lambda _{2} \\ 
\lambda _{0} & \lambda _{0} & \lambda _{1} & \lambda _{2} \\ 
\lambda _{1} & \lambda _{1} & \lambda _{2} & \lambda _{2} \\ 
\lambda _{2} & \lambda _{2} & \lambda _{2} & \lambda _{2}%
\end{array}%
.  \label{S2}
\end{equation}

Following the S-expansion method discussed in Appendices A and B to expand
the extended de Sitter algebra, $\mathcal{A}d\mathcal{S}^{E}$ (\ref{AdSE1})-(%
\ref{AdSE6}), using the semigroup (\ref{S2}), considering the partition in
subspaces (\ref{V0V1}) y (\ref{V0V1-2}), and the resonant partition (\ref%
{S0S1a}), we have the resonating subspaces (\ref{W0W1}) 
\begin{equation}
W_{0}=S_{0}\times V_{0}=\left \{ \lambda _{0}J^{\mu \nu },\lambda
_{0}L_{a}^{C_{1}...C_{n}},\lambda _{2}J^{\mu \nu },\lambda
_{2}L_{a}^{C_{1}...C_{n}}\right \}  \label{W0}
\end{equation}%
\begin{equation}
W_{1}=S_{1}\times V_{1}=\left \{ \lambda _{1}P^{\mu },\lambda _{2}P^{\mu
}\right \}  \label{W1}
\end{equation}%
and the resonant subalgebra (\ref{BR}), 
\begin{equation}
\mathcal{B}_{R}=W_{0}\oplus W_{1}=\left \{ \lambda _{0}J^{\mu \nu },\lambda
_{0}L_{a}^{C_{1}...C_{n}},\lambda _{2}J^{\mu \nu },\lambda
_{2}L_{a}^{C_{1}...C_{n}},\lambda _{1}P^{\mu },\lambda _{2}P^{\mu }\right \}
.  \label{WR}
\end{equation}

Choosing the zero element $\lambda _{2}=0_{S}$ of the semigroup, we obtain
the reduced resonant subalgebra (\ref{zero})%
\begin{equation}
\mathcal{B}_{R0}=\left \{ \lambda _{0}J^{\mu \nu },\lambda
_{0}L_{a}^{C_{1}...C_{n}},\lambda _{1}P^{\mu }\right \} .  \label{WR0}
\end{equation}

We redefine the previous generators in the form

\begin{equation}
\begin{array}{ccc}
\mathcal{J}^{\mu \nu }=\lambda _{0}J^{\mu \nu }, & l\mathcal{P}^{\mu
}=\lambda _{1}P^{\mu }, & \mathcal{L}_{a}^{C_{1}...C_{n}}=\lambda
_{0}L_{a}^{C_{1}...C_{n}},%
\end{array}
\label{RG}
\end{equation}%
using the extended de Sitter algebra, $\mathcal{A}d\mathcal{S}^{E}$ (\ref%
{AdSE1})-(\ref{AdSE6}), the semigroup (\ref{S2}), and extending the flat
limit $l\longrightarrow \infty $; we obtain the extended Poincaré algebra ($%
\mathcal{B}_{3}^{E}$), (\ref{PP})-(\ref{LL}), that corresponds to the same
result obtained in (9), (10) and (11) of reference \cite{Antoniadis2011}.

\subsection{Extended AdS$\oplus $Lorentz ($\mathcal{C}_{4}^{E}$) as a
resonant and reduced subalgebra of extended de Sitter algebra ($\mathcal{C}%
_{3}^{E}$).}

\ \ \ To obtain extended AdS$\oplus $Lorentz ($\mathcal{C}_{4}^{E}$) \cite%
{P. Salgado2014}, \cite{CR1}, \cite{CCFRS}, \cite{CCRS}, we should consider
the semigroup $S^{(4)}=\left \{ \lambda _{0},\lambda _{1},\lambda
_{2},\lambda _{3}\right \} $, whose multiplication table is the following 
\begin{equation}
S^{(4)}=%
\begin{array}{ccccc}
\ast & \lambda _{0} & \lambda _{1} & \lambda _{2} & \lambda _{3} \\ 
\lambda _{0} & \lambda _{0} & \lambda _{1} & \lambda _{2} & \lambda _{3} \\ 
\lambda _{1} & \lambda _{1} & \lambda _{2} & \lambda _{1} & \lambda _{3} \\ 
\lambda _{2} & \lambda _{2} & \lambda _{1} & \lambda _{2} & \lambda _{3} \\ 
\lambda _{3} & \lambda _{3} & \lambda _{3} & \lambda _{3} & \lambda _{3}%
\end{array}
\label{S3}
\end{equation}%
(note that the table (24) is embedding in the table (30)), with the
partition of the semigroup%
\begin{equation}
\begin{array}{ccc}
S_{0}=\left \{ \lambda _{0},\lambda _{2},\lambda _{3}\right \} & , & 
S_{1}=\left \{ \lambda _{1},\lambda _{3}\right \} ,%
\end{array}
\label{S0S1AdS}
\end{equation}%
which satisfies the resonant condition (\ref{CR}).

Following the S-expansion method \cite{Sexp}-\cite{PDE} of Appendices A and
B, using the subspaces (\ref{V0V1}), (\ref{V0V1-2}) and the decomposition (%
\ref{S0S1AdS}), we have the resonating subspaces (\ref{W0W1})%
\begin{equation}
W_{0}=S_{0}\times V_{0}=\left \{ \lambda _{0}J^{\mu \nu },\lambda
_{0}L_{a}^{C_{1}...C_{n}},\lambda _{2}J^{\mu \nu },\lambda
_{2}L_{a}^{C_{1}...C_{n}},\lambda _{3}J^{\mu \nu },\lambda
_{3}L_{a}^{C_{1}...C_{n}}\right \} ,  \label{W02}
\end{equation}%
and%
\begin{equation}
W_{1}=S_{1}\times V_{1}=\left \{ \lambda _{1}P^{\mu },\lambda _{3}P^{\mu
}\right \} ,  \label{W12}
\end{equation}%
also the corresponding resonant subalgebra (\ref{BR}), 
\begin{equation}
\mathcal{B}_{R}=W_{0}\oplus W_{1}=\left \{ 
\begin{array}{c}
\lambda _{0}J^{\mu \nu },\lambda _{0}L_{a}^{C_{1}...C_{n}},\lambda
_{2}J^{\mu \nu },\lambda _{2}L_{a}^{C_{1}...C_{n}}, \\ 
\lambda _{3}J^{\mu \nu },\lambda _{3}L_{a}^{C_{1}...C_{n}},\lambda
_{1}P^{\mu },\lambda _{3}P^{\mu }%
\end{array}%
\right \} .  \label{SAR2}
\end{equation}

Defining $\lambda _{3}=0_{S}$ as the zero of the semigroup, the resonant and
reduced subalgebra is the following%
\begin{equation}
\mathcal{B}_{R0}=W_{0}\oplus W_{1}=\left \{ \lambda _{0}J^{\mu \nu },\lambda
_{0}L_{a}^{C_{1}...C_{n}},\lambda _{2}J^{\mu \nu },\lambda
_{2}L_{a}^{C_{1}...C_{n}},\lambda _{1}P^{\mu }\right \} .  \label{SARR2}
\end{equation}

We must impose an extra condition between some elements of the semigroup,
for the algebra to be closed, that is to say,%
\begin{equation}
\lambda _{2}:=\lambda _{1},  \label{Constr}
\end{equation}%
therefore the semigroup (\ref{S3}) is reduced to the semigroup%
\begin{equation}
S_{red.}^{(4)}=%
\begin{array}{cccc}
\ast & \lambda _{0} & \lambda _{1} & \lambda _{3} \\ 
\lambda _{0} & \lambda _{0} & \lambda _{1} & \lambda _{3} \\ 
\lambda _{1} & \lambda _{1} & \lambda _{1} & \lambda _{3} \\ 
\lambda _{3} & \lambda _{3} & \lambda _{3} & \lambda _{3}%
\end{array}%
.  \label{S3R}
\end{equation}%
($\mathcal{S}_{red.}^{(4)}$ does not lose the associative property due to
the (\ref{Constr}) condition),

Taking into account the condition (\ref{Constr}), the semigroup (\ref{S3R})
continues to satisfy the resonant condition (\ref{CR}) and we define the
generators of (\ref{SARR2}) in the form

\begin{eqnarray}
&&%
\begin{array}{ccc}
\mathcal{J}^{\mu \nu }=\lambda _{0}J^{\mu \nu }, & \mathcal{P}^{\mu
}=\lambda _{1}P^{\mu }, & \mathcal{Z}^{\mu \nu }=\lambda _{1}J^{\mu \nu },%
\end{array}
\notag \\
&&%
\begin{array}{cc}
\mathcal{L}_{a0}^{C_{1}...C_{n}}=\lambda _{0}L_{a}^{C_{1}...C_{n}}, & 
\mathcal{L}_{a1}^{C_{1}...C_{n}}=\lambda _{1}L_{a}^{C_{1}...C_{n}}.%
\end{array}
\label{RDG}
\end{eqnarray}

Using the generators (\ref{RDG}) and the semigroup (\ref{S3R}), we get by
expansion the extension of the AdS$\oplus $Lorentz algebra ($\mathcal{C}%
_{4}^{E}$) from the extension of de Sitter algebra (\ref{AdSE1})-(\ref{AdSE6}%
):%
\begin{eqnarray}
\left[ \mathcal{P}^{\mu },\mathcal{P}^{\nu }\right] &=&-i\mathcal{Z}^{\mu
\nu },  \notag \\
\left[ \mathcal{J}^{\mu \nu },\mathcal{P}^{\rho }\right] &=&i(g^{\rho \nu }%
\mathcal{P}^{\mu }-g^{\rho \mu }\mathcal{P}^{\nu }),  \notag \\
\left[ \mathcal{J}^{\mu \nu },\mathcal{J}^{\rho \sigma }\right] &=&i(g^{\mu
\sigma }\mathcal{J}^{\nu \rho }-g^{\mu \rho }\mathcal{J}^{\nu \sigma
}+g^{\nu \rho }\mathcal{J}^{\mu \sigma }-g^{\nu \sigma }\mathcal{J}^{\mu
\rho }),  \notag \\
\left[ \mathcal{J}^{\mu \nu },\mathcal{Z}^{\rho \sigma }\right] &=&i(g^{\mu
\sigma }\mathcal{Z}^{\nu \rho }-g^{\mu \rho }\mathcal{Z}^{\nu \sigma
}+g^{\nu \rho }\mathcal{Z}^{\mu \sigma }-g^{\nu \sigma }\mathcal{Z}^{\mu
\rho }),  \label{AdSL} \\
\left[ \mathcal{Z}^{\mu \nu },\mathcal{P}^{\rho }\right] &=&i(g^{\rho \nu }%
\mathcal{P}^{\mu }-g^{\rho \mu }\mathcal{P}^{\nu }),  \notag \\
\left[ \mathcal{Z}^{\mu \nu },\mathcal{Z}^{\rho \sigma }\right] &=&i(g^{\mu
\sigma }\mathcal{Z}^{\nu \rho }-g^{\mu \rho }\mathcal{Z}^{\nu \sigma
}+g^{\nu \rho }\mathcal{Z}^{\mu \sigma }-g^{\nu \sigma }\mathcal{Z}^{\mu
\rho }),  \notag
\end{eqnarray}

\begin{eqnarray}
\left[ \mathcal{J}^{\mu \nu },\mathcal{L}_{ai}^{C_{1}...C_{s}}\right]
&=&i(\eta ^{C_{1}\nu }\mathcal{L}_{ai}^{\mu C_{2}...C_{s}}-...-\eta
^{C_{s}\mu }\mathcal{L}_{ai}^{C_{1}...C_{s-1}\nu }),  \notag \\
\left[ \mathcal{P}^{\mu },\mathcal{L}_{ai}^{C_{1}...C_{s}}\right] &=&i(\eta
^{C_{1}\mu }\mathcal{L}_{a1}^{4C_{2}...C_{s}}-...-\eta ^{C_{s}4}\mathcal{L}%
_{a1}^{C_{1}...C_{s-1}\mu }),\text{ }  \notag \\
\left[ \mathcal{Z}^{\mu \nu },\mathcal{L}_{ai}^{C_{1}...C_{s}}\right]
&=&i(\eta ^{C_{1}\nu }\mathcal{L}_{a1}^{\mu C_{2}...C_{s}}-...-\eta
^{C_{s}\mu }\mathcal{L}_{a1}^{C_{1}...C_{s-1}\nu }),  \label{AdSLE} \\
\left[ \mathcal{L}_{ai}^{C_{1}...C_{n}},\mathcal{L}_{bi}^{C_{n+1}...C_{s}}%
\right] &=&if_{abc}\mathcal{L}_{ci}^{C_{1}...C_{s}},  \notag \\
\left[ \mathcal{L}_{a0}^{C_{1}...C_{n}},\mathcal{L}_{b1}^{C_{n+1}...C_{s}}%
\right] &=&if_{abc}\mathcal{L}_{c1}^{C_{1}...C_{s}},  \notag \\
&&%
\begin{array}{cccc}
where & s=0,1,2,..... & and & i=0,1%
\end{array}
\notag
\end{eqnarray}

Note that in this case the generators $\left \{ 
\begin{array}{cc}
\mathcal{L}_{ai}^{C_{1}...C_{s}}, & i=0,1%
\end{array}%
\right \} $ carry higher-spins gauge fields and differently from the case of
the Poincaré algebra, here the higher-spins generators are not
translationally invariant, because the commutation relation (\ref{AdSE5}) is
not null. The algebra (\ref{AdSL})-(\ref{AdSLE}) satisfy all Jacoby
identities (\ref{JC}) because it is an expansion (see apppendix A) of the
extended algebra (\ref{AdSE1})-(\ref{AdSE6}) and the latter is only obtained
by fixing some indexes of extended de Sitter algebra (\ref{SLE1})-(\ref{SLE4}%
).

The first six commutators (\ref{AdSL}) correspond to the AdS$\oplus $Lorentz
algebra and the remaining (\ref{AdSLE}) correspond to the extended sector.
Note that if we do the identification 
\begin{equation*}
\mathcal{J}^{\mu \nu }\longleftrightarrow \mathcal{Z}^{\mu \nu }
\end{equation*}%
the extended $AdS$ algebra as subalgebra of extended AdS$\oplus $Lorentz, is
obtained. \ The last five commutations relations describe the coupling of
the higher-spin generators $\left \{ \mathcal{L}_{ai}^{C_{1}...C_{s}}\text{, 
}i=0,1\right \} $ to the AdS$\oplus $Lorentz symmetry.

\subsection{The extended Maxwell algebra ($\mathcal{B}_{4}^{E}$) as a flat
limit of extended AdS$\oplus $Lorentz ($\mathcal{C}_{4}^{E}$) algebra.}

\ \ \ To obtain the extended Maxwell algebra ($\mathcal{B}_{4}^{E}$), we
must introduce the $l$ parameter in the definition of the generators in (\ref%
{RDG}), in the form

\begin{eqnarray}
&&%
\begin{array}{ccc}
\mathcal{J}^{\mu \nu }=\lambda _{0}J^{\mu \nu }, & l\mathcal{P}^{\mu
}=\lambda _{1}P^{\mu }, & l^{2}\mathcal{Z}^{\mu }=\lambda _{1}J^{\mu \nu },%
\end{array}
\notag \\
&&%
\begin{array}{cc}
\mathcal{L}_{a0}^{C_{1}...C_{n}}=\lambda _{0}L_{a}^{C_{1}...C_{n}}, & l%
\mathcal{L}_{a1}^{C_{1}...C_{n}}=\lambda _{1}L_{a}^{C_{1}...C_{n}}.%
\end{array}
\label{RDG2}
\end{eqnarray}

Using the extended de Sitter algebra ($\mathcal{C}_{3}^{E}$) (\ref{AdSE1})-(%
\ref{AdSE6}), the semigroup (\ref{S3R}), and by extending the singular limit 
$l\longrightarrow \infty $ (flat limit), we obtain by expansion the
extension of the Maxwell algebra ($\mathcal{B}_{4}^{E}$) from the resonant
and reduced subalgebra (\ref{SARR2})%
\begin{eqnarray}
\left[ \mathcal{P}^{\mu },\mathcal{P}^{\nu }\right] &=&-i\mathcal{Z}^{\mu
\nu },  \notag \\
\left[ \mathcal{J}^{\mu \nu },\mathcal{P}^{\rho }\right] &=&i(g^{\rho \nu }%
\mathcal{P}^{\mu }-g^{\rho \mu }\mathcal{P}^{\nu }),  \notag \\
\left[ \mathcal{J}^{\mu \nu },\mathcal{J}^{\rho \sigma }\right] &=&i(g^{\mu
\sigma }\mathcal{J}^{\nu \rho }-g^{\mu \rho }\mathcal{J}^{\nu \sigma
}+g^{\nu \rho }\mathcal{J}^{\mu \sigma }-g^{\nu \sigma }\mathcal{J}^{\mu
\rho }),  \notag \\
\left[ \mathcal{J}^{\mu \nu },\mathcal{Z}^{\rho \sigma }\right] &=&i(g^{\mu
\sigma }\mathcal{Z}^{\nu \rho }-g^{\mu \rho }\mathcal{Z}^{\nu \sigma
}+g^{\nu \rho }\mathcal{Z}^{\mu \sigma }-g^{\nu \sigma }\mathcal{Z}^{\mu
\rho }),  \label{RCB4A} \\
\left[ \mathcal{Z}^{\mu \nu },\mathcal{P}^{\rho }\right] &=&0,  \notag \\
\left[ \mathcal{Z}^{\mu \nu },\mathcal{Z}^{\rho \sigma }\right] &=&0,  \notag
\end{eqnarray}
\begin{eqnarray}
\left[ \mathcal{J}^{\mu \nu },\mathcal{L}_{ai}^{C_{1}...C_{s}}\right]
&=&i(\eta ^{C_{1}\nu }\mathcal{L}_{ai}^{\mu C_{2}...C_{s}}-...-\eta
^{C_{s}\mu }\mathcal{L}_{ai}^{C_{1}...C_{s-1}\nu }),  \notag \\
\left[ \mathcal{P}^{\mu },\mathcal{L}_{ai}^{C_{1}...C_{s}}\right] &=&0,\text{
}  \notag \\
\left[ \mathcal{Z}^{\mu \nu },\mathcal{L}_{ai}^{C_{1}...C_{s}}\right] &=&0, 
\notag \\
\left[ \mathcal{L}_{a0}^{C_{1}...C_{n}},\mathcal{L}_{b0}^{C_{n+1}...C_{s}}%
\right] &=&if_{abc}\mathcal{L}_{c0}^{C_{1}...C_{s}},  \label{RCB4E} \\
\left[ \mathcal{L}_{a0}^{C_{1}...C_{n}},\mathcal{L}_{b1}^{C_{n+1}...C_{s}}%
\right] &=&if_{abc}\mathcal{L}_{c1}^{C_{1}...C_{s}},  \notag \\
\left[ \mathcal{L}_{a1}^{C_{1}...C_{n}},\mathcal{L}_{b1}^{C_{n+1}...C_{s}}%
\right] &=&0,  \notag \\
&&%
\begin{array}{cccc}
where & s=0,1,2,..... & and & i=0,1%
\end{array}
\notag
\end{eqnarray}

Analogously, the first six commutators (\ref{RCB4A}) correspond to the
Maxwell algebra and the remaining (\ref{RCB4E}) correspond to the extended
sector that describe the coupling of the higher-spin generators $\left \{ 
\mathcal{L}_{ai}^{C_{1}...C_{s}}\text{, }i=0,1\right \} $ to the Maxwell
symmetry.

\subsection{Extended $\mathcal{C}_{k}^{E}$ algebra as a resonant and reduced
subalgebra of extended de Sitter algebra ($\mathcal{C}_{3}^{E}$).}

\ \ \ Following the same methodology; semigroups $S^{(5)}$, $S^{(6)}$, $%
S^{(7)},...$were obtained, where it was possible to see explicitly that each
semigroup was embedded in the next of higher order, in the same way as the
semigroup $S^{(2)}$ (\ref{S2}) is embedded in $S^{(3)}$ (\ref{S3}). The
semigroups $S^{(5)}$, $S^{(6)}$, $S^{(7)},...$ were obtained using the
general multiplication rule shown below in $i$, $ii$, $iii$, $iv$. \ The
following Lie algebras: $\mathcal{C}_{5}^{E}$ , $\mathcal{C}_{6}^{E}$ , $%
\mathcal{C}_{7}^{E},...$ and its corresponding flat limits ($%
l\longrightarrow \infty $) $\mathcal{B}_{5}^{E}$ , $\mathcal{B}_{6}^{E}$ , $%
\mathcal{B}_{7}^{E},...$ can be obtained directly using the S-expansion
method \cite{Sexp}-\cite{PDE} (appendices A and B) and the corresponding
semigroups $S^{(5)}$, $S^{(6)},S^{(7)},...$

Following this inductive method, the generalized extended $\mathcal{C}%
_{k}^{E}$ algebras can be obtained, using the general semigroup 
\begin{equation}
S^{(k)}=\left \{ \lambda _{0},\lambda _{1},\lambda _{2},\lambda
_{3},...,\lambda _{k-1}\right \} ,
\end{equation}%
with the resonant partition

\begin{equation}
\begin{array}{ccc}
S_{0}=\left \{ \lambda _{0},\lambda _{2},...,\lambda _{2n},...,\lambda
_{k-1}\right \} & , & S_{1}=\left \{ \lambda _{1},\lambda _{3},...,\lambda
_{2n+1},...,\lambda _{k-1}\right \}%
\end{array}%
,  \label{S0S1G}
\end{equation}%
and generalizing the multiplication rules used in (\ref{S2}) and (\ref{S3}).

Keeping in mind that $\lambda _{0}$ is the unit element (monoid structure),
where 
\begin{equation}
\lambda _{0}\ast \lambda _{n}=\lambda _{n}\text{,}
\end{equation}%
for all $\lambda _{n}$ $\in S^{(k)}$, we have generalized the multiplication
rules in (\ref{S2}) and (\ref{S3}), as follows:

$%
\begin{array}{c}
\\ 
\end{array}%
$

$i)$ Lambda with even index, squared:%
\begin{equation}
\lambda _{2n}\ast \lambda _{2n}:=\left( \lambda _{2n}\right) ^{2}=\lambda
_{2n}  \label{par2}
\end{equation}%
where $n=0,1,2,...$

$ii)$ Lambda with even index by Lambda with even index: 
\begin{equation}
\begin{array}{ccc}
\lambda _{2n}\ast \lambda _{2m}=\lambda _{2m}\ast \lambda _{2n}=\lambda _{2}
& where & n,m\neq 0%
\end{array}
\label{parpar}
\end{equation}

$iii)$ Lambda with odd index by Lambda with odd index: 
\begin{equation}
\begin{array}{ccccc}
\lambda _{2n+1}\ast \lambda _{2m+1}=\lambda _{2} & when & n\neq m & and & n=m%
\end{array}
\label{imparimpar}
\end{equation}%
where $n,m=0,1,2,...$

$iv)$ Lambda with even index by Lambda with odd index: 
\begin{equation}
\begin{array}{ccccc}
\lambda _{2n}\ast \lambda _{2m+1}=\lambda _{2m+1}\ast \lambda _{2n}=\lambda
_{1} & where & n\neq 0 & y & m=0,1,2,...%
\end{array}
\label{parimpar}
\end{equation}%
where $\lambda _{2n}$ , $\lambda _{2n+1}\in S^{(k)}$ and $n=0,1,2,...$

It is straightforward to verify that multiplication rules $i$, $ii$, $iii$
and $iv$ reproduce cases (\ref{S2}), (\ref{S3}) and also \ the cases $%
S^{(5)} $, $S^{(6)}$ and $S^{(7)}.$

Using the partition (\ref{S0S1G}), and applying the S-expansion procedure
shown in Appendices A and B (\ref{W0W1}), \cite{Sexp}, \cite{IPRS}, we have
the resonating subspaces%
\begin{eqnarray*}
W_{0} &=&\left \{ 
\begin{array}{c}
\lambda _{0}J^{\mu \nu },\lambda _{0}L_{a}^{C_{1}...C_{n}},\lambda
_{2}J^{\mu \nu },\lambda _{2}L_{a}^{C_{1}...C_{n}},...,\lambda _{2n}J^{\mu
\nu },\lambda _{2n}L_{a}^{C_{1}...C_{n}}, \\ 
...,\lambda _{k-1}J^{\mu \nu },\lambda _{k-1}L_{a}^{C_{1}...C_{n}}%
\end{array}%
\right \} \\
&& \\
W_{1} &=&\left \{ \lambda _{1}P^{\mu },\lambda _{3}P^{\mu },...,\lambda
_{2n+1}P^{\mu },...,\lambda _{k-1}P^{\mu }\right \}
\end{eqnarray*}%
and the resonant subalgebra 
\begin{equation}
\mathcal{B}_{R}=W_{0}\oplus W_{1}=\left \{ 
\begin{array}{c}
\lambda _{0}J^{\mu \nu },\lambda _{0}L_{a}^{C_{1}...C_{n}},\lambda
_{2}J^{\mu \nu },\lambda _{2}L_{a}^{C_{1}...C_{n}},... \\ 
,\lambda _{2n}J^{\mu \nu },\lambda _{2n}L_{a}^{C_{1}...C_{n}}, \\ 
...,\lambda _{k-1}J^{\mu \nu },\lambda _{k-1}L_{a}^{C_{1}...C_{n}}, \\ 
\\ 
\lambda _{1}P^{\mu },\lambda _{3}P^{\mu },...,\lambda _{2n+1}P^{\mu
},...,\lambda _{k-1}P^{\mu }.%
\end{array}%
\right \} .  \label{WRF}
\end{equation}

Choosing the zero of the semigroup $\lambda _{k-1}=0$, we obtain the reduced
resonant subalgebra 
\begin{equation}
\mathcal{B}_{R0}=W_{0}\oplus W_{1}=\left \{ 
\begin{array}{c}
\lambda _{0}J^{\mu \nu },\lambda _{0}L_{a}^{C_{1}...C_{n}},\lambda
_{2}J^{\mu \nu },\lambda _{2}L_{a}^{C_{1}...C_{n}},... \\ 
,\lambda _{2n}J^{\mu \nu },\lambda _{2n}L_{a}^{C_{1}...C_{n}},... \\ 
,\lambda _{1}P^{\mu },\lambda _{3}P^{\mu },...,\lambda _{2n+1}P^{\mu },...%
\end{array}%
\right \} .
\end{equation}

Knowing that, the multiplication rules of $ii$ and $iv$ are compatible with
the condition (\ref{Constr}), without entering into contradiction we can
generalize it, as follows

\begin{equation}
\begin{array}{ccc}
\lambda _{2n}=\lambda _{2n-1} & where & n=1,2,...%
\end{array}%
,  \label{Cond.gen}
\end{equation}%
where the case $n=1$ and $k=4$ reproduces the condition (\ref{Constr}) and
the semigroup (\ref{S3}).

We can define the generators as follows 
\begin{equation}
\begin{array}{ccc}
\mathcal{J}_{2n}^{\mu \nu }=\lambda _{2n}J^{\mu \nu }, & \mathcal{P}%
_{2m+1}^{\mu }=\lambda _{2m+1}P^{\mu }, & \mathcal{L}_{a2n}^{C_{1}...C_{s}}=%
\lambda _{2n}L_{a}^{C_{1}...C_{n}}%
\end{array}
\label{GCk}
\end{equation}%
where $n=0,1,2,..$.

Using the generators (\ref{GCk}) and (\ref{RDG}), the multiplication rules(%
\ref{par2}), (\ref{parpar}), (\ref{imparimpar}) and (\ref{parimpar}), the
extended $AdS^{E}$ algebra (\ref{AdSE1})-(\ref{AdSE6}), and the general
condition (\ref{Cond.gen}); the commutations relations of extended $\mathcal{%
C}_{k}^{E}$ algebra, are 
\begin{equation}
\left[ \mathcal{J}_{2n}^{\mu \nu },\mathcal{J}_{2m}^{\mu \nu }\right]
=i(g^{\mu \sigma }\mathcal{Z}^{\nu \rho }-g^{\mu \rho }\mathcal{Z}^{\nu
\sigma }+g^{\nu \rho }\mathcal{Z}^{\mu \sigma }-g^{\nu \sigma }\mathcal{Z}%
^{\mu \rho }),  \label{Ck1}
\end{equation}%
\begin{equation}
\left[ \mathcal{J}_{2n}^{\mu \nu },\mathcal{P}_{2m+1}^{\rho }\right]
=i(g^{\rho \nu }\mathcal{P}^{\mu }-g^{\rho \mu }\mathcal{P}^{\nu }),
\label{Ck2}
\end{equation}%
\begin{equation}
\left[ \mathcal{J}_{2n}^{\mu \nu },\mathcal{L}_{a2m}^{C_{1}...C_{s}}\right]
=i(\eta ^{C_{1}\nu }\mathcal{L}_{a1}^{\mu C_{2}...C_{s}}-...-\eta ^{C_{s}\mu
}\mathcal{L}_{a1}^{C_{1}...C_{s-1}\nu }),  \label{Ck3}
\end{equation}%
\begin{equation}
\left[ \mathcal{P}_{2n+1}^{\mu },\mathcal{P}_{2m+1}^{\nu }\right] =-i%
\mathcal{Z}^{\mu \nu },  \label{Ck4}
\end{equation}%
\begin{equation}
\left[ \mathcal{P}_{2n+1}^{\mu },\mathcal{L}_{a2m}^{C_{1}...C_{s}}\right]
=i(\eta ^{C_{1}\mu }\mathcal{L}_{a1}^{4C_{2}...C_{s}}-...-\eta ^{C_{s}4}%
\mathcal{L}_{a1}^{C_{1}...C_{s-1}\mu }),  \label{Ck5}
\end{equation}%
\begin{equation}
\left[ \mathcal{L}_{a2n}^{C_{1}...C_{s}},\mathcal{L}_{a2m}^{C_{1}...C_{s}}%
\right] =if_{abc}\mathcal{L}_{c1}^{C_{1}...C_{s}},  \label{Ck6}
\end{equation}

Note that when $n=0,1$ and $m=0,1$ the generators of (\ref{RDG}) are
rescued, where $\mathcal{J}_{0}^{\mu \nu }\rightarrow \mathcal{J}^{\mu \nu }$
and $\mathcal{J}_{2}^{\mu \nu }\rightarrow \mathcal{Z}^{\mu \nu }$\ and we
obtain the already known commutations relations of extended AdS$\oplus $%
Lorentz ($\mathcal{C}_{4}^{E}$) algebra (\ref{AdSL}), (\ref{AdSLE}), where $%
\mathcal{C}_{4}^{E}\subseteq $ $\mathcal{C}_{k}^{E}$.\bigskip \ \ 

\subsection{Extended $\mathcal{B}_{k}^{E}$ algebra as a flat limit of
extended $\mathcal{C}_{k}^{E}$ algebra.}

\ \ \ The extended $\mathcal{B}_{k}^{E}$ algebras are obtained as the flat
limit of the extended $\mathcal{C}_{k}^{E}$ algebras, when $l\longrightarrow
\infty .$ In effect, introducing the $l$ parameter in the definition of the
generators (\ref{GCk}) in the form 
\begin{eqnarray}
&&%
\begin{array}{cc}
....,l^{2n}\mathcal{J}_{2n}^{\mu }=\lambda _{2n}J^{\mu \nu },...., & l^{2m+1}%
\mathcal{P}_{2m+1}^{\mu }=\lambda _{2m+1}P^{\mu },%
\end{array}
\label{RGCk} \\
&&%
\begin{array}{ccc}
& l^{2n}\mathcal{L}_{a2n}^{C_{1}...C_{s}}=\lambda _{2n}L_{a}^{C_{1}...C_{n}},
& 
\end{array}
\notag
\end{eqnarray}%
the computing of the commutation relations of the generators (\ref{RGCk})
and by extending the singular limit $l\longrightarrow \infty $, we obtain
the commutation relations of extended and generalized $\mathcal{B}_{k}^{E}$
algebra, which corresponds to the flat limit of the commutation relations of
the extended $\mathcal{C}_{k}^{E}$ algebra (\ref{Ck1})-(\ref{Ck6}):

$\circ $ For $n>1$ and $m>0$ 
\begin{equation}
\left[ \mathcal{J}_{2n}^{\mu \nu },\mathcal{J}_{2m}^{\mu \nu }\right] =0,
\end{equation}%
\begin{equation}
\left[ \mathcal{J}_{2n}^{\mu \nu },\mathcal{P}_{2m+1}^{\rho }\right] =0,
\end{equation}%
\begin{equation}
\left[ \mathcal{J}_{2n}^{\mu \nu },\mathcal{L}_{a2m}^{C_{1}...C_{s}}\right]
=0,
\end{equation}%
\begin{equation}
\left[ \mathcal{P}_{2n+1}^{\mu },\mathcal{P}_{2m+1}^{\nu }\right] =0,
\end{equation}%
\begin{equation}
\left[ \mathcal{P}_{2n+1}^{\mu },\mathcal{L}_{a2m}^{C_{1}...C_{s}}\right] =0,
\end{equation}%
\begin{equation}
\begin{array}{cc}
\left[ \mathcal{L}_{a2n}^{C_{1}...C_{s}},\mathcal{L}_{a2m}^{C_{1}...C_{s}}%
\right] =0, & s=0,1,2,...%
\end{array}%
\end{equation}%
\ \ 

$\circ $ For $n=0,1$ and $m=0$ is reduced to the already known commutation
relations of extended Maxwell algebra ($\mathcal{B}_{4}^{E}\subseteq 
\mathcal{B}_{k}^{E}$) (\ref{RCB4A})-(\ref{RCB4E}), where $\mathcal{J}%
_{0}^{\mu \nu }=\lambda _{0}\mathcal{J}^{\mu \nu }$, $l\mathcal{P}_{1}^{\rho
}=\lambda _{1}P^{\mu }$, $l^{2}\mathcal{J}_{2}^{\mu }=\lambda _{1}J^{\mu \nu
}$ and $\mathcal{J}_{2}^{\mu \nu }\rightarrow \mathcal{Z}^{\mu \nu }$ of (%
\ref{RDG2}).

\section{Comments and possible developments}

\ \ \ Starting from higher spin extended de Sitter and Conformal Lie algebra
($\mathcal{C}_{3}^{E}$) proposed by \cite{Antoniadis2011}, \cite%
{Antoniadis2012} we can recover all the other families of higher spin
extended $\mathcal{C}_{k}^{E}$ and its flat limit, the $\mathcal{B}_{k}^{E}$
algebras. Specifically, from extended de Sitter algebra ($\mathcal{C}%
_{3}^{E} $) and using the S-expansion method of Lie algebra (shown in
Appendices A and B), we have obtained as a resonant and reduced subalgebras:
the extended Poincaré algebra ($\mathcal{B}_{3}^{E}$), the extended AdS$%
\oplus $Lorentz algebra ($\mathcal{C}_{4}^{E})$ and its flat limit the
extended Maxwell algebra ($\mathcal{B}_{4}^{E}$), and their generalizations,
the extended $\mathcal{C}_{k}^{E}$ algebra and its corresponding flat limit
the extended $\mathcal{B}_{k}^{E}$ algebra \cite{P. Salgado2014}, \cite{CR1}%
. The above results are summarized in the following diagram, i.e,

\begin{equation*}
\begin{tabular}{ccccc}
\cline{1-1}\cline{3-3}\cline{5-5}
\multicolumn{1}{|c}{$\overset{}{\mathcal{C}_{4}^{E}}$} & \multicolumn{1}{|c}{%
$\overset{{\small S-}\exp {\tiny .}}{\longleftarrow }$} & 
\multicolumn{1}{|c}{$\mathcal{C}_{3}^{E}$} & \multicolumn{1}{|c}{$...\overset%
{S-\exp .}{\longrightarrow }$} & \multicolumn{1}{|c|}{$\overset{}{\mathcal{C}%
_{k}^{E}}$} \\ \cline{1-1}\cline{3-3}\cline{5-5}
&  &  & $\ $ &  \\ 
\ \ \ \ \ $\downarrow $ ${\tiny (l\rightarrow \infty )}$ &  & $\downarrow $ $%
{\tiny (l\rightarrow \infty )}$ &  & $\ \ \ \ \downarrow $ ${\tiny %
(l\rightarrow \infty )}$ \\ 
&  &  & $\ $ &  \\ \cline{1-1}\cline{3-3}\cline{5-5}
\multicolumn{1}{|c}{$\overset{}{\mathcal{B}_{4}^{E}}$} & \multicolumn{1}{|c}{
} & \multicolumn{1}{|c}{$\mathcal{B}_{3}^{E}$} & \multicolumn{1}{|c}{$...$}
& \multicolumn{1}{|c|}{$\overset{}{\mathcal{B}_{k}^{E}}$} \\ 
\cline{1-1}\cline{3-3}\cline{5-5}
\end{tabular}%
\end{equation*}%
\begin{equation*}
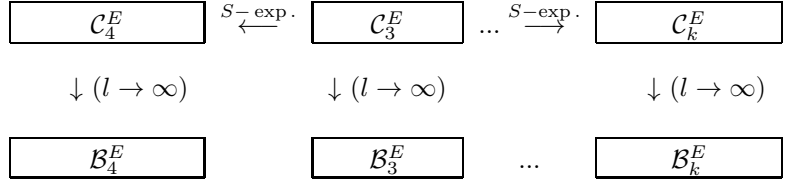
\text{Figure 1: Map between different extended algebras and their
relationships. }
\end{equation*}

\ As possible developments we will investigate the irreducible
representations of the extended Maxwell algebra ($\mathcal{B}_{4}^{E}$) and
extended $\mathcal{B}_{k}^{E}$ family algebras for massless cases. Both the
longitudinal and transversal representations would be studied, such as the
case of extended (Super)Poincaré algebras in Ref. \cite{Antoniadis2011}, 
\cite{Antoniadis2012}. Continuous Spin Representations of the Poincaré and
Super-Poincaré Groups has been built in Ref. \cite{BKRX}, \cite{EW}. \
Another possible developments, we will intend to study the expansions of
(Super)symmetric version of extended de Sitter algebra in four space-time
dimensions. \bigskip

\section{ Acknowledgment}

\ \ \ This work was supported by the Research Project, Code DIN 11/2012 of
the Universidad Católica de la Santísima Concepción, Chile. R.C. would like
to thank to P. Salgado and S. Salgado for his useful comments.

\section*{Appendix}

\appendix

\section{Brief introduction to the S-expansion method}

\ \ \ Given a finite abelian semigroup $S=\left \{ \lambda _{0},\lambda
_{1},...,\lambda _{N+1}\right \} $, with a commutative and associative
composition law $S\times S\rightarrow S$,\bigskip 
\begin{equation}
\begin{array}{ccc}
\lambda _{\alpha }\lambda _{\beta }=K_{\alpha \beta }^{\gamma }\lambda
_{\gamma } & , & \alpha =0,1,...,N+1%
\end{array}
\label{ec1.}
\end{equation}%
and be the pair $(\mathcal{G};[;])$ a Lie algebra, where $\mathcal{G}$ is a
finite dimensional vector space, with basis $\left \{ T_{A}\right \}
_{A=1}^{\dim G}$ over the field $K$ of the real or complex numbers; and $[;]$
is a rule of composition $\mathcal{G\times G\rightarrow G}$,%
\begin{equation}
\left( T_{A},T_{B}\right) \rightarrow \left[ T_{A},T_{B}\right]
=C_{AB}^{C}T_{C}.  \label{ec2}
\end{equation}

The direct product $\mathcal{B=}S\times \mathcal{G}$ is defined as 
\begin{equation}
\mathcal{B=}\left \{ 
\begin{array}{ccccc}
T_{\left( A,\alpha \right) }=\lambda _{\alpha }T_{A} & : & \lambda _{\alpha }
& \in S, & T_{A}\in G%
\end{array}%
\right \}  \label{ec3}
\end{equation}%
provided with the composition law $[;]_{S}$: $\mathcal{B\times B\rightarrow B%
}$ defined by%
\begin{equation}
\left[ T_{\left( A,\alpha \right) },T_{\left( B,\beta \right) }\right]
=\lambda _{\alpha }\lambda _{\beta }\left[ T_{A},T_{B}\right] =K_{\alpha
\beta }^{\gamma }C_{AB}^{C}\lambda _{\gamma }T_{C}=C_{\left( A,\alpha
\right) \left( B,\beta \right) }^{\left( C,\gamma \right) }T_{\left(
C,\gamma \right) }  \label{ec4}
\end{equation}%
where 
\begin{equation}
C_{\left( A,\alpha \right) \left( B,\beta \right) }^{\left( C,\gamma \right)
}=K_{\alpha \beta }^{\gamma }C_{AB}^{C}.  \label{SC}
\end{equation}

The prior equation defines the Lie bracket of the S-expanded Lie algebra,
where $T_{\left( A,\alpha \right) }=\lambda _{\alpha }T_{A}$ is a base of $%
\mathcal{B}$, with the composition law (\ref{ec4}) and it is called a
S-expanded Lie algebra \cite{Sexp}-\cite{PDE}. \ The expanded algebra (\ref%
{ec3}) satisfy the linearity, antisymmetry and the Jacobi identity. Indeed,
the expanded structure constants satisfy the Jacobi condition, that is to
say,%
\begin{equation}
\frac{1}{2}\varepsilon _{ABC}^{DEF}C_{\left( D,\alpha \right) \left( E,\beta
\right) }^{\left( G,\delta \right) }C_{\left( G,\delta \right) \left(
F,\gamma \right) }^{\left( H,\zeta \right) }=0.  \label{JC}
\end{equation}

\section{Resonant ($\mathcal{B}_{R}$) and reduced subalgebras ($\mathcal{B}%
_{R0}$) of $\mathcal{B=}S\times \mathcal{G}$}

\ \ \ According to the S-expansion method, a decomposition of the abelian
semigroup $S$ can be make, $S=S_{0}\cup S_{1}$ \cite{Sexp}, \cite{IPRS},
when the algebra $\mathcal{G}$ can be splitting as a direct sum of a
subalgebra and symmetric coset, $\mathcal{G=}V_{0}\oplus V_{1}$, that is to
say%
\begin{equation}
\begin{array}{ccccc}
\left[ V_{0},V_{0}\right] \subset V_{0} & , & \left[ V_{0},V_{1}\right]
\subset V_{1} & , & \left[ V_{1},V_{1}\right] \subset V_{0},%
\end{array}
\label{V0V1-2}
\end{equation}%
and it is said that such decomposition is resonant, when the following
conditions are satisfied%
\begin{equation}
\begin{array}{ccccc}
S_{0}\times S_{0}\subset S_{0} & , & S_{0}\times S_{1}\subset S_{1} & , & 
S_{1}\times S_{1}\subset S_{0}.%
\end{array}
\label{CR}
\end{equation}

The resonant subalgebras $\mathcal{B}_{R}\subseteq \mathcal{B}$ can be
obtained by construction 
\begin{equation}
\mathcal{B}_{R}=W_{0}\oplus W_{1},  \label{BR}
\end{equation}%
where 
\begin{equation}
\begin{array}{cccc}
W_{0}=S_{0}\otimes V_{0} & and & W_{1}=S_{1}\otimes V_{1} & ,%
\end{array}
\label{W0W1}
\end{equation}%
and reduced resonant subalgebras, $\mathcal{B}_{R0}\subseteq \mathcal{B}%
_{R}\subseteq \mathcal{B}$, can be obtained by choosing a zero element of
the semigroup, $\lambda _{N+1}:=0_{S}$, and therefore 
\begin{equation}
T_{(A,N+1)}=\lambda _{N+1}T_{A}=0.  \label{zero}
\end{equation}

This causes some of commutation relations to be null (abelianize) and the
process to be equivalent with an Inönü-Wigner contraction.

\emph{Therefore, the resonant and reduced subalgebras are not mere copies of
the original algebra, but are more complex structures.}

For the case when the $\mathcal{G}$ algebra presents a structure of
superalgebra; that is to say%
\begin{equation}
\mathcal{G=}V_{0}\oplus V_{1}\oplus V_{2}  \label{SA}
\end{equation}%
where 
\begin{eqnarray}
&&%
\begin{array}{ccc}
\left[ V_{0},V_{0}\right] \subset V_{0}, & \left[ V_{0},V_{1}\right] \subset
V_{1}, & \left[ V_{0},V_{2}\right] \subset V_{2}%
\end{array}
\label{CSA} \\
&&%
\begin{array}{ccc}
\left[ V_{1},V_{1}\right] \subset V_{0}\oplus V_{2}, & \left[ V_{1},V_{2}%
\right] \subset V_{1}, & \left[ V_{2},V_{2}\right] \subset V_{0}\oplus V_{2},%
\end{array}
\notag
\end{eqnarray}%
in analogy to the previous case, we can make a resonant decomposition $%
S=S_{0}\cup S_{1}\cup S_{2}$, which satisfy the following conditions%
\begin{align}
& 
\begin{array}{ccc}
S_{0}\times S_{0}\subset S_{0}, & S_{0}\times S_{1}\subset S_{1}, & 
S_{0}\times S_{2}\subset S_{2}%
\end{array}
\label{CRSA} \\
& 
\begin{array}{ccc}
S_{1}\times S_{1}\subset S_{0}\cup S_{2}, & S_{1}\times S_{2}\subset S_{1},
& S_{2}\times S_{2}\subset S_{0}\cup S_{2}.%
\end{array}
\notag
\end{align}

Following the S-expansion procedure, we can build the resonant subalgebras 
\begin{equation}
\mathcal{B}_{R}=W_{0}\oplus W_{1}\oplus W_{2},  \label{BR2}
\end{equation}%
where%
\begin{equation}
\begin{array}{ccccccc}
W_{0}=S_{0}\otimes V_{0} & , & W_{1}=S_{1}\otimes V_{1} & , & 
W_{2}=S_{2}\otimes V_{2} & , & 
\end{array}
\label{W0W1W2}
\end{equation}%
and the reduced resonant subalgebras $\mathcal{B}_{R0}$, choosing a zero
element of the semigroup, e.g. $\lambda _{N+1}:=0_{S}$.\bigskip \

\end{document}